\documentclass[10pt,a4paper,aps,twocolumn,showpacs,amssymb,amsmath, color,floatfix]{revtex4}
\usepackage[utf8x]{inputenc}
\usepackage[english]{babel}
\usepackage[T2A]{fontenc}
\usepackage{ucs}
\usepackage{amsmath}
\usepackage{amsfonts}
\usepackage{amssymb}
\usepackage{graphicx}
\graphicspath{{EPS_figs/}{PDF_figs/}}
\usepackage{color}
\usepackage{bm}

\begin{document}

\title{Elastic octopoles and colloidal structures in nematic liquid crystals}
\author{S. B. Chernyshuk $^{1)}$, O.M. Tovkach $^{2)}$ and B. I. Lev $^{2)}$}
\affiliation{$^{1)}$ Institute of Physics, NAS Ukraine, Prospekt
Nauki 46, Kyiv 03650, Ukraine}
\affiliation{ $^{2)}$ Bogolyubov Institute of
Theoretical Physics, NAS Ukraine, Metrologichna 14-b, Kyiv 03680,
Ukraine.}

\begin{abstract}
We propose a simple theoretical model which explains a formation of dipolar 2D and 3D colloidal structures in nematic liquid crystal. Colloidal particles are treated as effective hard spheres interacting via their elastic dipole, quadrupole and octopole moments. It is shown that octopole moment plays an important role in the formation of 2D and 3D nematic colloidal crystals. We generalize this assumption on the case of the external electric field and theoretically explain a giant electrostriction effect in 3D crystals observed recently [A. Nych et al., Nature Communications \textbf{4}, 1489 (2013)].
\end{abstract}

\maketitle

Nematic liquid crystal (NLC) colloids have attracted significant research interest during the last decades. Particles, suspended in a liquid crystal host, cause director field distortions which give rise to a new class of elastic interactions. These long-range anisotropic interactions result in different colloidal structures: 1D linear chains \cite{po1,po2}, inclined chains with respect to the director  \cite{po3}-\cite{lavr1} and 2D nematic colloidal crystals \cite{nych}-\cite{ulyana}. Recently a 3D colloidal crystal was experimentally observed for the first time \cite{Nych_2013}. 

Small director deformations as well as electric field potential are governed by the Laplace equation. Thus theoretical understanding of the elastic interactions is based on the multipole expansion of the director field deformations and has deep electrostatic analogies. Untill now axially symmetrical particles were considered to have only dipole and quadrupole elastic terms \cite{lupe}-\cite{we4}, assuming higher order elastic terms to be neglected. 

At the same time high order electric moments play an important role in different areas of physics. For instance octopole moment has a significant importance in nuclear physics and in intermolecular interactions. For example the methane molecule $CH_{4}$ has zero dipole and quadrupole moments and nonzero octopole electric moment \cite{Kaplan}. In general any pear-like charge distribution has nonzero octopole moment. 

Hedgehog director configuration as well has pear-like form, so it's natural that octopole elastic moment should be manifested in the elastic colloidal interactions. In this Letter we show that this is a truth.  

Let's now consider an axially symmetrical particle in the NLC. The immersed particle induces deformations of the director in the perpendicular directions $n_{\mu}, \mu=x,y$ and make director field $\textbf{n}\approx(n_{x},n_{y},1)$. The bulk energy of deformation may be approximately written in the harmonic form:
\begin{equation}
F_{har}=\frac{K}{2}\int  d^{3}x (\nabla n_{\mu})^{2}\label{fb3}
\end{equation}
with Euler-Lagrange equations of Laplace type:
\begin{equation}
\Delta n_{\mu}=0  \label{lap}
\end{equation}
Then the director field outside the particle in the simplest case has the form  $n_{x}(\textbf{r})=p\frac{x}{r^{3}}+3c\frac{xz}{r^{5}},n_{y}(\textbf{r})=p\frac{y}{r^{3}}+3c\frac{yz}{r^{5}}$  with $p$ and $c$ being dipole and quadrupole elastic moments. The anharmonic correction to the bulk energy is $F_{anhar}=\frac{K}{2}\int  d^{3}x (\nabla n_{z})^{2}\approx\frac{K}{8}\int  d^{3}x (\nabla n_{\bot}^{2})^{2}$ which changes EL equations to be:
\begin{equation}
\Delta n_{\mu}+\frac{1}{2}n_{\mu}\Delta n_{\bot}^{2}=0  \label{unhar}
\end{equation}
If the leading contribution to $n_{\mu}$ is the dipolar term then anharmoic corrections are of the form $r_{\mu}/r^{7}$ and high order terms of the order up to $1/r^{5}$ can effectively influence on the short-range behaviour and should be equally considered. 

In the general case, the solution of the Laplace equation for axially symmetric particles has the form: 
\begin{equation}
n_{\mu}=\sum^{N}_{l=1}a_{l}(-1)^{l}\partial_{\mu}\partial_{z}^{l-1}\frac{1}{r}  \label{hn}
\end{equation}
where $a_{l}$ is the multipole moment of the order $l$ and $2^{l}$ is the multipolarity; $N$ - is the maximum possible order without anharmonic corrections. For the dipole particle $N=4$. So $a_{1}=p$ is the dipole moment, $a_{2}=c$ is the quadrupole moment, $a_{3}$ is the octopole moment, $a_{4}$ is the hexadecapole moment.  

 In order to find the energy of the system: particle(s) + LC , it is necessary to introduce some effective free energy functional $F_{eff}$ so that it's Euler-Lagrange equations would have the above solutions (\ref{hn}). In the one constant approximation with Frank constant $K$ the effective functional has the form:
\begin{equation}
F_{eff}=K\int d^{3}x\left\{\frac{(\nabla n_{\mu})^{2}}{2}-4\pi\sum^{N}_{l=1}A_{l}(\textbf{x})\partial_{\mu}\partial_{z}^{l-1}n_{\mu} \right\}\label{flin}
\end{equation}
which brings Euler-Lagrange equations:
\begin{equation}
\Delta n_{\mu}=4\pi\sum^{N}_{l=1}(-1)^{l-1}\partial_{\mu}\partial_{z}^{l-1}A_{l}(\textbf{x})\label{nmu}
\end{equation}
where $A_{l}(\textbf{x})$ are multipole moment densities, $\mu=x,y$ and repeated $\mu$ means summation on $x$ and $y$ like $\partial_{\mu}n_{\mu}=\partial_{x}n_{x}+\partial_{y}n_{y}$.
For the bulk NLC the solution has the known form: 
\begin{equation}
n_{\mu}(\textbf{x})=\int d^{3}\textbf{x}' \frac{1}{\left|\textbf{x}-\textbf{x}'\right|}\sum^{N}_{l=1}(-1)^{l}\partial_{\mu}'\partial_{z}'^{l-1}A_{l}(\textbf{x}') \label{solmain} 
\end{equation}
If we consider $A_{l}(\textbf{x})=a_{l}\delta(\textbf{x})$ this really brings solution (\ref{hn}). This means that effective functional (\ref{flin}) correctly describes the interaction between the particle and LC. 

 Consider $N_{p}$ particles in the NLC, so that $A_{l}(\textbf{x})=\sum_{i}a_{l}^{i}\delta(\textbf{x}-\textbf{x}_{i})$, $i=1\div N_{p}$ . Then substitution (\ref{solmain}) into $F_{eff}$ (\ref{flin}) brings: $F_{eff}=U^{self}+U^{interaction}$ where $U^{self}=\sum_{i}U_{i}^{self} $ , here $U_{i}^{self}$ is the divergent self energy.\\
Interaction energy $ U^{interaction}=\sum_{i<j}U_{ij} $.  Here $U_{ij}$ is the elastic interaction energy between $i$ and $j$ particles in the bulk NLC: 
\begin{equation}
U_{ij}=4\pi K \sum^{N}_{l,l'=1}a_{l}a_{l'}'(-1)^{l'}\frac{(l+l')!}{r^{l+l'+1}}P_{l+l'}(cos\theta)\label{uint}
\end{equation}
Here unprimed quantities $a_{l}$ are used for particle $i$ and primed $a_{l'}'$ for particle $j$, $r=|\textbf{x}_{i}-\textbf{x}_{j}|$, $\theta$ is the angle between $\textbf{r}$ and z and we used the relation $P_{l}(cos\theta)=(-1)^{l}\frac{r^{l+1}}{l!}\partial_{z}^{l}\frac{1}{r}$ for Legendre polynomials $P_{l}$. It is the general expression for the elastic interaction potential between axially symmetric colloidal particles in the bulk NLC with taking into account of the high order elastic terms. In what follows below for dipole particles we suppose $a_{4}=0$ and $N=3$, so that particles have nonzero dipole, quadrupole and octopole moments ($a_{1},a_{2},a_{3}$).

This formula was first obtained in \cite{high}, where it was used for the description of the interaction between beads with planar anchoring and boojums director configuration. It was found there that $a_{4}$ and $a_{6}$ moments ($N=6$ for quadrupole particles) give the angle $\theta_{min}=34.5^{\circ}$ between two contact beads which is close to the experimental value of $\theta_{min}=30^{\circ}$ \cite{po3}. 

\begin{figure}
\begin{center}
\includegraphics[width=6cm]{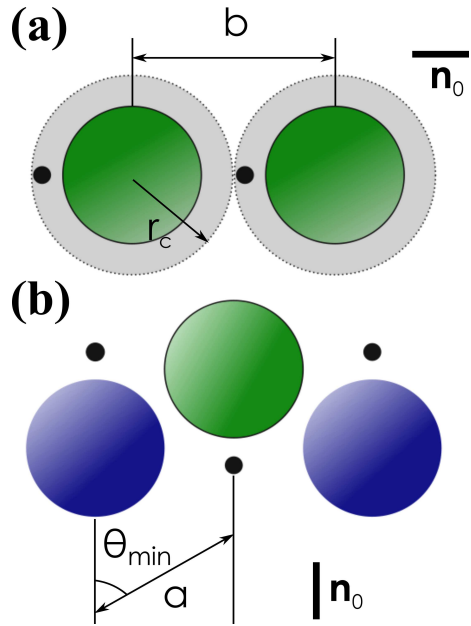}
\end{center}
\caption{ \textbf{(a)} 1D colloidal structure. Particles with parallel dipole moments aggregate in linear chains along $\mathbf{n}_0$, $b \approx 2.4r_0$ \cite{Fukuda_2004, Noel_2006}. Grey zone is the \textit{coat}. \textbf{(b)} A zigzag vertical cross-section of a quasi-2D checkerboard colloidal crystal formed by particles with antiparallel hedgehogs ordering in the homeotropic cell, $a \approx 2.3r_0$ and $\theta_{\text{min}} \approx 60^{\circ}$ \cite{Nych_2013}.
Each particle is surrounded by the \textit{coat} containing strong director deformations which cannot be described by the multipole expansion.}\label{1D}
\end{figure}

Of course there is always the nearest zone, where  formula \eqref{uint} is not applicapable \cite{high}. This is the \textit{coat} zone (see Fig.\ref{1D}a), where topological defects are concentrated and anharmonic terms are essential. The average equilibrium distance $b$ between the centers of the dipole particles in the chain (taken from different experiments as well as numerical calulations \cite{Noel_2006, Fukuda_2004, Nych_2013, Mus} ) is $b=2.44r_{0}$. This means that we can take radius of the coat to be $r_c = 1.22 r_0$ and suppose that the short-range part of the interaction potential is close enough to the potential of hard spheres. So that the total effective interaction potential between two dipole particles has the form:

\begin{equation}\label{U_full_oct}
\frac{U}{4 \pi K} = \begin{cases}
\sum_{l,l^{\prime}=1}^{3} a_{l} a_{l^{\prime}}^{\prime} (-1)^{l^{\prime}} \dfrac{(l+l^{\prime})!}{r^{l+l^{\prime}+1}} P_{l+l^{\prime}}(\cos\theta)\,\, , &r > 2 r_c \\
\infty\,\, , &r \le 2 r_c 
\end{cases}
\end{equation} 
where $a_1 = \alpha r_0^2, a_2 = -\beta r_0^3, a_3 = \gamma r_0^4$ are dipole, quadrupole and octopole elastic moments, respectively.

Suppose first that octopole moment is zero $a_{3}=0$. 

In the paper \cite{Noel_2006} authors used iron particles with dipole director configuration and made precise direct measurements of the elastic forces due to the balance between the elastic and magnetic forces in the equilibrium position. They found $\alpha_{exp} = 2.05$ and $\beta_{exp} = 0.2 \pm 0.1$. At the same time in \cite{lupe} it was theoretically found from the special dipole ansatz that  $\alpha_{theor} = 2.04$ and $\beta_{theor} = 0.72$. Here is very good correspondence for the dipole moment and so bad for the quadrupole moment, though the same value $\beta_{theor} = 0.72$ was obtained for two different ansatzes in \cite{lupe}. Why?

\begin{figure}
\begin{center}
\includegraphics[width=\columnwidth]{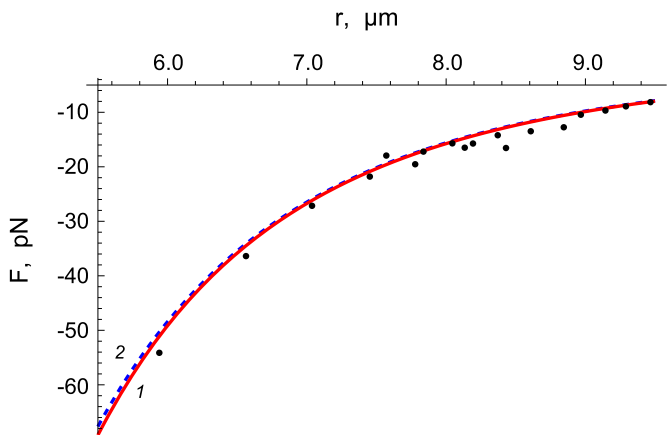}
\end{center}
\caption{The attractive part of the elastic force between two parallel dipoles. Points depict the experimental results from \cite{Noel_2006}. Solid red line 1 corresponds to the case $\alpha_{exp} = 2.05$, $\beta_{exp} = 0.2$ and $\gamma = 0$. Dashed blue line 2 corresponds to the following coefficients $\alpha_{theor} = 2.04$, $\beta_{theor} = 0.72$ and $\gamma = 0.157$.}\label{Noel}
\end{figure}
We think that the reason is the neglect of the octopole moment. The dipole-octopole interaction is exactly the same as the quadrupole-quadrupole interaction $U_{QQ}+U_{dO}=(a_{2}a_{2}'-a_{1}a_{3}'-a_{3}a_{1}')\frac{24P_{4}(\cos\theta)}{r^{5}}$ for axially symmetric particles. If we suppose that authors of \cite{lupe} correctly found the quadrupole moment $\beta = 0.72$, then we can estimate the octopole moment from the comparison with results of \cite{Noel_2006}: $\beta_{exp}^{2}\approx\beta_{theor}^{2}-2\alpha\gamma$ so that $\gamma\approx 0.12$. More precisely we can fit Noel's results with expression \eqref{U_full_oct} and easily find that $\gamma = 0.157$. Herewith the difference between two curves, $(\alpha,\beta,\gamma)=(2.05, 0.2, 0)$ and $(2.04, 0.72, 0.157)$, is lower than 0.3\% for all the experimental points (see Fig.\ref{Noel}).

\begin{figure}
\begin{center}
\includegraphics[width=6cm]{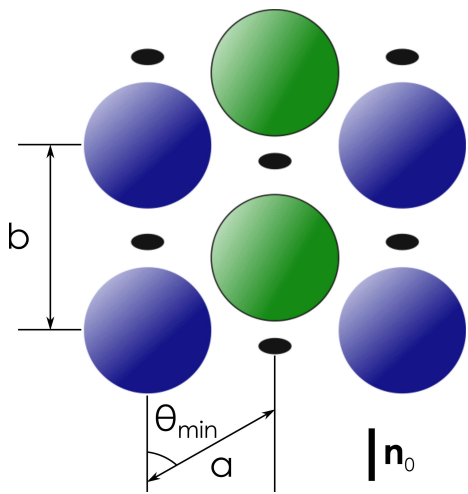}
\end{center}
\caption{2D structure formed by antiparallel linear chains in the planar cell. The lattice constants are $a = 2.54r_0$, $b = 2.46r_0$ and $\theta_{\text{min}} = 61^{\circ}$ \cite{Mus}. Note that in such a structure the hedgehogs transform into small rings.}\label{2D}
\end{figure}

Now let us consider antiparallel dipoles in the homeotropic cell ($a_1=-a_1^{\prime}$ and $a_{2}=a_{2}'$) . As it was reported in \cite{Nych_2013} such particles form a quasi-2D checkerboard colloidal crystal. Fluorescent confocal polarizing microscopy (FCPM) provides a vertical cross-section which has a zigzag form with interparticle distance $a \approx 2.3r_0$ and azimuthal angle $\theta_{\text{min}} \approx 60^{\circ}$ (see Fig.\ref{1D}b). To explain this structure we must minimize energy \eqref{U_full_oct} over two variables: $r$ and $\theta$. Simple calculation for $(\alpha,\beta,\gamma)=(2.05,0.2,0)$ gives  $\theta_{\text{min}} = 83.3^{\circ}$. Obviously this value is far from reality.

Repeating the same for antiparallel dipoles with $(\alpha,\beta,\gamma)=(2.04,0.72,0.157)$ (note that here $a_{1} = -a_{1}'$, $a_{2} = a_{2}'$ and $a_{3} = -a_{3}'$) it is easy to ensure that $\theta_{\text{min}} = 63^{\circ}$, which is consistent with experimental value $60^{\circ}$ \cite{Nych_2013}.

In addition to this 2D structure, a 2D hexagonal crystal formed by antiparallel dipolar chains in the planar cell has been observed (see Fig.\ref{2D}). Treating the energy of such a system as the sum of pair energies \eqref{U_full_oct} with $(\alpha,\beta,\gamma)=(2.04,0.72,0.157)$ we can find the lattice parameters: $a = b = 2.44r_0$ and $\theta_{\text{min}} = 64.2^{\circ}$. These parameters are in agreement with experimental values $a = 2.54r_0$, $b = 2.46r_0$ and $\theta_{\text{min}} = 61^{\circ}$ \cite{Mus}.

\begin{figure}
\begin{center}
\includegraphics[width=6cm]{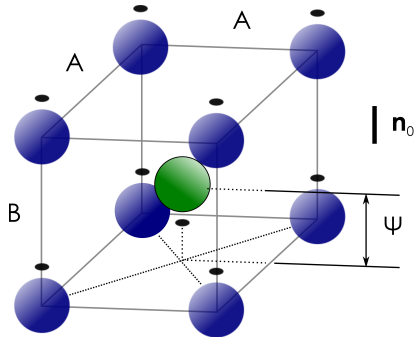}
\end{center}
\caption{Sketch of a quasi body-centred Bravais lattice of a 3D colloidal crystal. The lattice constants are $A = (3.2 \pm 0.1)r_0$, $B = (2.3 \pm 0.2)r_0$ and $\Psi = (1.3 \pm 0.1)r_0$ \cite{Nych_2013}. Due to the dense packing the central particle interacts only with its nearest neighbours located at the vertices.}\label{3D}
\end{figure}
 
Recently Nych \textit{et al.} first reported about experimental observation of a 3D colloidal crystals with the tetragonal symmetry (see Fig.\ref{3D}). The lattice constants were recorded directly from the FCPM images and found to be $A_{exp} = (3.2 \pm 0.1)r_0$, $B_{exp} = (2.3 \pm 0.2)r_0$ and $\Psi_{exp} = (1.3 \pm 0.1)r_0$ \cite{Nych_2013}. To simplify our calculations we suppose that in such a dense-packed structure every particle interacts only with its nearest neighbours. Then again minimizing the energy of the lattice for parameters $(\alpha,\beta,\gamma)=(2.04,0.72,0.157)$ we find $A_{theor} = 3.07r_0$, $B_{theor} = 2.44r_0$ and $\Psi_{theor} = 1.1r_0$ . We use the multipole coefficients for the hyperbolic hedgehog configuration, but in the 2D and 3D structures the hedgehogs open up into small rings and probably this can alter the coefficients $(\alpha,\beta,\gamma)$ a little bit.

\begin{figure}
\begin{center}
\includegraphics[width=\columnwidth]{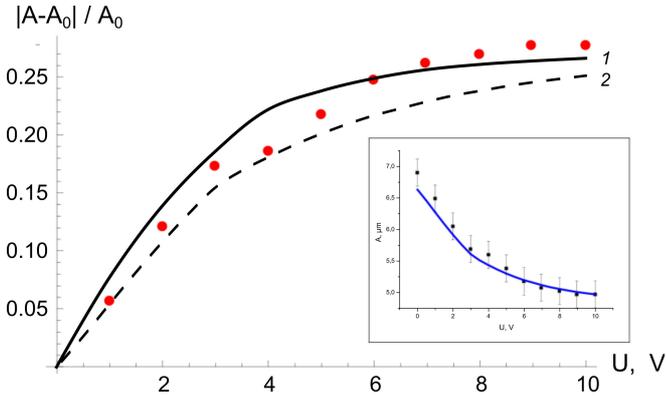}
\end{center}
\caption{Lateral shrinking of a 3D colloidal crystal under an electric field applied along $\mathbf{n}_0$. Theoretical curves were obtained with the following parameters of E7: dielectric anisotropy $\Delta\epsilon = 13.8$, elastic constant $K =13.7 \text{pN}$, the cell thickness $L = 25 \mu$m, $\eta = 1.6$ (solid line 1) and $\eta = 1.0$ (dashed line 2). Inset shows the lattice "width" $A$ as a function of applied voltage for $4.32 \mu$m particles and $\eta = 1.0$. Points depict experimental results from \cite{Nych_2013}.}\label{Shrinking}
\end{figure}

The 3D colloidal crystal in the NLC with positive dielectric anisotropy exhibits the, so called, giant electrostriction, i.e. lateral shrinking under the action of the electric field applied along $\mathbf{n}_0$ \cite{Nych_2013}. The influence of the electric field on the colloidal interactions has been early discussed in \cite{we3}. There it was shown that in a bulk nematic with $\Delta\epsilon > 0$ the field gives rise to the exponential screening of the multipole interactions. The energy of the pair interaction in this case has the form \cite{we3,high}:
\begin{equation}\label{U_full_field}
\frac{U}{4 \pi K} = \begin{cases}
-\sum_{l,l^{\prime}=1}^{3} a_{l} a_{l^{\prime}}^{\prime} \partial_{\mu}\partial_{\mu}^{\prime} \partial_{z}^{l-1}\partial_{z}^{\prime l^{\prime}-1} G(\mathbf{x}, \mathbf{x}^{\prime})\,\, , &r > 2 r_c \\
\infty\,\, , &r \le 2 r_c 
\end{cases}
\end{equation}
where $\partial_{\mu}\partial_{\mu}^{\prime} = \partial_{x}\partial_{x}^{\prime} + \partial_{y}\partial_{y}^{\prime}$ , $G(\mathbf{x}, \mathbf{x}^{\prime}) = \exp\left[ -|\mathbf{x} - \mathbf{x}^{\prime}|/\xi \right]/|\mathbf{x} - \mathbf{x}^{\prime}|$ is the Green's function for a bulk nematic with the electric field and $\xi = \frac{1}{E}\sqrt{\frac{K}{4\pi\epsilon_0\Delta\epsilon}}$ is the electric coherence length.  
Obviously, in some way the field should affect on the particles coats as well so that $r_c=r_c(E)$. The simplest assumption that we can make about it is the following. If $\Delta\epsilon > 0$, the nematic molecules have a tendency to align along the field direction. Thus we have a competition between this aligning and the anchoring on the particle surface. Apparently, the further from the surface we are and the smaller the elastic constant $K$ is, the easier molecules can be reoriented. The same in the language of mathematics
\begin{equation}\label{r_c}
-\frac{dx}{dE} \propto \frac{x}{K},
\end{equation}
where $x = r_c -r_0$. From the dimensional analysis it follows that \eqref{r_c} can be rewritten as
\begin{equation}\label{r_c_full}
-\frac{dx}{x(E)} = \eta r_0 \sqrt{\frac{4\pi\epsilon_0 \Delta\epsilon}{K}} dE
\end{equation}
where $\eta$ is some dimensionless parameter. And finally taking into account that $x(E=0)=0.22r_0$ we arrive at
\begin{equation}
r_c = r_0 \left(1+0.22 e^{-\frac{\eta r_0}{\xi}}\right)
\end{equation}
It is well known \cite{stark1} that a transition from the hedgehog to the Saturn-ring occurs at the field strength $r_0/\xi = 3.3$ . In the experiments we are talking about $r_0/\xi < 3.3$ and the symmetry of the director field remains dipolar. Due to this we can assume that the particle's coat does not shrink along the defect axis (along $\mathbf{n}_0$). This means that the coats, in fact, are not the spheres but are rather prolate spheroids and the lattice "height", $B$, practically does not depend on $E$. 
Taking this into account and minimizing the energy of the lattice over $A$ and $\Psi$ one can find the lattice "width", $A$, as a function of the field strength. The results of these calculations are shown in Fig.\ref{Shrinking}. The only fitting parameter here is $\eta$ (all other parameters are known). We see that the correspondence is rather good that as well confirms the importance of the octopole moment. 

In conclusion, we have shown that the elastic octopole moment plays an important role in the formation of 2D and 3D dipolar colloidal crystals. It is found that the elastic octopole moment $a_{3}=\gamma r_{0}^4$ in the hedgehog director configuration has the approximate value $\gamma=0.157$. This value can explain the characteristics of all the dipolar 2D and 3D colloidal structures treating the colloids as the effective hard spheroids interacting via their elastic dipole, quadrupole and octopole moments. 

Generalization of this idea on the case of the electric field presence can explain the effect of the giant electrostriction observed recently.


\begin{thebibliography}{99}

\bibitem{po1}  P.Poulin, H.Stark, T.C.Lubensky and D.A.Weitz, Science \textbf{275}, 1770 (1997).
\bibitem{po2}  P.Poulin1, V. Cabuil and D. A. Weitz ,  Phys.Rev. Lett. \textbf{79}, 4862 (1997).
\bibitem{po3}  P.Poulin and D.A.Weitz,  Phys.Rev. E \textbf{57}, 626 (1998).
\bibitem{lavr1} I.I.Smalyukh, O.D.Lavrentovich, A.N.Kuzmin, A.V.Kachynski and  P.N.Prasad, Phys. Rev. Lett. \textbf{95}, 157801 (2005)
\bibitem{nych}  V.Nazarenko, A.Nych and B.Lev, Phys.Rev.Lett. \textbf{87}, 075504 (2001).
\bibitem{R10} I. I. Smalyukh, S. Chernyshuk, B. I. Lev, A. B. Nych, U.Ognysta, V.G. Nazarenko, and O. D. Lavrentovich, Phys. Rev. Lett.
{\bf 93}, 117801 (2004).
\bibitem{Mus} I. Mu$\check{s}$evic, M. $\check{S}$karabot, U.Tkalec, M.Ravnik and S.$\check{Z}$umer Science  {\bf 313}, 954 (2006).
\bibitem{s1} M.$\check{S}$karabot, M. Ravnik, S.$\check{Z}$umer, U. Tkalec, I. Poberaj, D. Babi$\check{c}$, N. Osterman and I. Mu$\check{s}$evic, Phys.Rev.E \textbf{77}, 031705 (2008)
\bibitem{ulyana} U.Ognysta, A. Nych, V. Nazarenko, I. Mu$\check{s}$evic, M.$\check{S}$karabot, M. Ravnik, S.$\check{Z}$umer, I. Poberaj and D. Babi$\check{c}$,   
Phys.Rev.Lett. \textbf{100}, 217803 (2007)
\bibitem{Nych_2013} A. Nych, U. Ognysta, M. $\check{S}$karabot, M.Ravnik, S.$\check{Z}$umer and I. Mu$\check{s}$evic, Nature Communications \textbf{4}, 1489 (2013).

\bibitem{lupe} T.C.Lubensky, D.Pettey, N.Currier and H.Stark, Phys.Rev.E \textbf{57}, 610 (1998).
\bibitem{ram}  S. Ramaswamy, R. Nityananda, V. A. Gaghunathan, and J. Prost, Mol. Cryst. Liq. Cryst. \textbf{288}, 175 (1996).
\bibitem{lev}  B.I.Lev and P.M.Tomchuk,  Phys.Rev.E \textbf{59}, 591 (1999).
\bibitem{lev3} B.I.Lev, S.B.Chernyshuk, P.M.Tomchuk and H.Yokoyama, Phys.Rev.E \textbf{65}, 021709 (2002)
\bibitem{perg3} V. M. Pergamenshchik and V. A. Uzunova, Phys. Rev. E \textbf{83}, 021701 (2011)
\bibitem{we}  S.B.Chernyshuk and B.I.Lev, Phys.Rev. E \textbf{81}, 041701 (2010)
\bibitem{we2} S.B.Chernyshuk and B.I.Lev, Phys.Rev. E \textbf{84}, 011707 (2011)
\bibitem{we3} S. B. Chernyshuk, O. M. Tovkach and B. I. Lev, Phys. Rev. E  \textbf{85}, 011706 (2012).
\bibitem{we4} O. M. Tovkach, S. B. Chernyshuk and B. I. Lev, Phys. Rev. E   \textbf{86}, 061703 (2012).
\bibitem{Kaplan}  Kaplan I. G. Intermolecular Interactions: Physical Picture, Computational Methods, and Model Potentials (Wiley, 2006).
\bibitem{high} S. B. Chernyshuk, http://arxiv.org/pdf/1205.0218.pdf,  (submitted to EPJE).


\bibitem{Noel_2006}  C. M. Noel, G. Bossis, A.-M. Chaze, F. Giulieri and S. Lacis, Phys. Rev. Lett. \textbf{96}, 217801 (2006).

\bibitem{Fukuda_2004}  J. Fukuda, H. Stark, M. Yoneya and H. Yokoyama, Phys. Rev. E \textbf{69}, 041706 (2004).  

\bibitem{stark1}  H.Stark, Eur. Phys. J.B. \textbf{10}, 311 (1999)



\end{thebibliography}
\end{document}